\documentclass[twoside,leqno,twocolumn]{article}
\usepackage{ltexpprt}

\usepackage{cite,bm}
\usepackage{caption}
\usepackage{graphicx}
\usepackage{subfigure}
\usepackage{float}
\usepackage{enumerate}
\usepackage{empheq}
\usepackage{amsmath}
\usepackage{amssymb}
\usepackage{amsfonts}
\usepackage{textcomp}
\usepackage{tabularx}
\usepackage{booktabs}
\usepackage{mathtools}
\usepackage{multirow}

\usepackage{array}
\usepackage[table]{xcolor}

\usepackage[hyphens]{url}
\usepackage{cite}
\usepackage{color}
\usepackage{caption}
\usepackage{dsfont}
\usepackage[flushleft]{threeparttable}

\newcommand{\cmmnt}[1]{}
\begin{document}

\title{\Large  Physics Guided RNNs for Modeling Dynamical Systems: A Case Study in Simulating Lake Temperature Profiles}
\author{Xiaowei Jia$^{1}$\thanks{These authors have equal contribution.}, Jared Willard$^{1*}$, Anuj Karpatne$^2$, Jordan Read$^3$, \\ Jacob Zwart$^3$, Michael Steinbach$^1$, Vipin Kumar$^1$\\
\small\baselineskip=9pt $^1$Department of Computer Science and Engineering, University of Minnesota \\
\small\baselineskip=9pt $^2$Department of Computer Science and Engineering, Virginia Tech \\
\small\baselineskip=9pt $^3$U.S. Geological Survey\\
\small  $^1$\{jiaxx221,willa099,stei0062,kumar001\}@umn.edu, $^2$karpatne@vt.edu, $^3$\{jread,jzwart\}@usgs.gov
}
\date{}

\maketitle 

% Default Copyright Statement
\fancyfoot[R]{\scriptsize{Copyright \textcopyright\ 2019 by SIAM\\
Unauthorized reproduction of this article is prohibited}}
\begin{abstract} \small\baselineskip=9pt 
This paper proposes a physics-guided recurrent neural network model (PGRNN) that combines RNNs and physics-based models to leverage their complementary strengths and improve the modeling of physical processes. Specifically, we show that a PGRNN can improve prediction accuracy over that of physical models, while generating outputs consistent with physical laws, and achieving good generalizability. Standard RNNs, even when producing superior prediction accuracy, often produce physically inconsistent results and lack generalizability.  We further enhance this approach by using a pre-training method that leverages the simulated data from a physics-based model to address the scarcity of observed data.  %The PGRNN has the flexibility to incorporate additional physical constraints and we incorporate a density-depth relationship. 
%Both enhancements further improve PGRNN performance.  
Although we present and evaluate this methodology in the context of modeling the dynamics of temperature in lakes, it is applicable more widely to a range of scientific and engineering disciplines where mechanistic (also known as process-based) models are used, e.g., power engineering, climate science, materials science, computational chemistry, and biomedicine.

\end{abstract}

\section{Introduction}

Physics-based models of dynamical systems are often used to study engineering and environmental systems. Despite their extensive use, these models have several well-known limitations due to simplified representations of the physical processes being modeled or challenges in selecting appropriate parameters. Given rapid data growth due to advances in sensor technologies, there is a tremendous opportunity to systematically advance modeling in these domains by using machine learning (ML) methods. However, capturing this opportunity is contingent on a paradigm shift in data-intensive scientific discovery since the “black box” use of ML often leads to serious false discoveries in scientific applications~\cite{lazer2014parable,karpatne2017theory}. %(cite our TGDS paper and Google flu trend paper). 
This paper presents a novel methodology for combining physics-based models with state-of-the-art deep learning methods to leverage their complementary strengths. Although we present and evaluate this methodology in the context of modeling the dynamics of temperature in lakes, it is applicable more widely to a range of scientific and engineering disciplines where mechanistic (also known as process-based) models are used, e.g., power engineering, climate science, %weather forecasting, 
materials science, computational chemistry, and biomedicine.

Even though physics-based models are based on known laws that govern relationships between input and output variables, they often rely on a large number of unknown parameters. These parameters must be estimated (or calibrated) from observed data, which is often scarce. A standard approach for model calibration is to intelligently search the space of parameter combinations and choose parameter combinations that result in the best performance on training data. This approach is computationally expensive as well as highly prone to over-fitting. Another limitation is that the majority of physics-based models implement approximate forms of physical relationships, either due to incomplete knowledge of certain processes or for practical computing purposes. Such approximations introduce additional parameters to these models that must be calibrated from data, making the process of model calibration even more challenging. These issues individually or in combination can introduce biases and thus result in unacceptable performance. The limitations of physics-based models cut across discipline boundaries and are well known in the scientific community~\cite{lall2014debates}. %, gupta2014debates, mcdonnell2014debates}. %In fact, a top journal in hydrology recently published a series of papers discussing limitations of traditional physics-based hydrologic models.%~\cite{gupta2014debates}. 

ML models, which have found tremendous success in several commercial applications where large-scale data is available, e.g., computer vision, and natural language processing, are increasingly being considered as promising alternatives to physics-based models by the scientific community. However, direct application of black-box ML models to a scientific problem encounters three major challenges:
1. State of the art (SOA) ML models that are powerful enough to effectively represent spatial and temporal processes inherent in physical systems can often perform better than traditional empirical models (e.g., regression-based models) used by the science communities as an alternative to physics-based models. However, they require a lot of training data, which is scarce in most practical settings.
2.  Empirical models (including the SOA ML models) simply identify statistical relations between inputs and the system variables of interest (e.g., the temperature profile of the lake) without taking in to account any physical laws (e.g., conservation of energy or mass) and thus can produce results that are inconsistent with physical laws.
3. Relationships produced by empirical models can at best be valid only for the set of forcing variable combinations present in the training data and are unable to generalize to scenarios unseen in the training data. For example, a ML model trained on a water body for today's climate may not be accurate for future warmer climate scenarios. 
As an alternative approach to both physics-based and empirical models, we present Physics-Guided Recurrent Neural Network models (PGRNN) as a general framework for modeling physical phenomena with potential applications for many disciplines. PGRNN  incorporates physics into the ML model by generalizing the loss function to include physical laws as the third component beyond the traditional notions of error and model complexity~\cite{karpatne2017theory}. %(refer to our TGDS paper).

Our proposed Physics-Guided Recurrent Neural Networks  model (PGRNN) is developed in the context of lake water temperature modeling at a daily scale. The temperature of water in a lake is known to be an ecological “master factor”~\cite{magnuson1979temperature} that controls the growth, survival, and reproduction of fish~\cite{roberts2013fragmentation}. Warming water temperatures can increase the occurrence of aquatic invasive species~\cite{rahel2008assessing,roberts2017nonnative}, which may displace fish and native aquatic organisms, and result in more harmful algal blooms (HABs)~\cite{harris2017predicting}.  %,paerl2008blooms}. 
Understanding temperature change and the resulting biotic “winners and losers” is timely science that can also be directly applied to inform priority action for natural resources.

The PGRNN model has a number of novel aspects.  This model contains two parallel recurrent structures - a standard RNN flow and an energy flow to be able to capture the variation of energy balance over time. While the standard RNN flow models the temporal dependencies that better fit observed data, the energy flow aims to regularize the temporal progression of the model in a physically consistent fashion.

To further improve the learning performance with the scarcity of observed data, we propose a pre-training method that utilizes the simulated data generated by physics-based models. While the simulated data are not an accurate reflection of the observed data, this pre-training algorithm has the potential to produce a better initialized status for the learning model and thus requires less observed data to fine-tune model parameters. 
Finally, we show that the proposed PGRNN model has the flexibility to incorporate additional physical constraints that are involved in specific applications. For example, in the lake temperature simulation problem, predicted values of the temperature at different depths should be such that denser water is at a lower depth, which is known as the density-depth constraint~\cite{karpatne2017physics}. 

We evaluate the proposed method in a reasonably large real-world system, Lake Mendota in Wisconsin. This lake is chosen for evaluation, as it is one of the most extensively studied lake systems and plenty of observed data is available to evaluate the performance of any new approach. We show that the modeling of energy conservation can successfully improve the learning performance and the generalization capacity. Moreover, the results confirm that the pre-training method can help achieve a reasonable performance even with a small amount of observed data. Finally, we show that after incorporating the density-depth constraint, the PGRNN model can produce both highly accurate and physically meaningful predictions.

\section{Problem Formulation}
%Since observational data of water temperature at
%broad spatial scales is incomplete (or non-existent in
%some regions) high-quality temperature modeling is
%necessary. Of particular interest is the problem of
To fully capture the temperature change in a lake system, we are interested in simulating the temperature of water at each depth $d$, and on each date, $t$. %This problem is referred to
%as 1D-modeling of temperature (depth being the single
%dimension). 
% A number of physics-based models have been developed for studying lake temperature, e.g., the
% state-of-the-art general lake model (GLM) [7]. 
% This model captures a variety of physical processes governing
% the dynamics of temperature in a lake, e.g., the heating
% of the water surface due to incoming shortwave radiation
% from the sun, the attenuation of radiation beneath the
% surface and the mixing of layers with varying energies
% at different depths, and the dissipation of heat from
% the surface of the lake via evaporation or longwave
% radiation, shown pictorially in Figure 3. We use GLM
% as our preferred choice of physics-based model for lake
% temperature modeling

Specifically, we consider the physical variables governing the dynamics of lake
temperature at every depth and time-step as the set of input drivers, $X=\{x_{d,t}\}$. These chosen features are known to be the primary drivers of lake thermodynamics \cite{hipsey2017general}. This includes meteorological recordings at the surface of water such as the amount of solar radiation, wind speed, air temperature, etc. %, as well as the value of depth. 
Given the input drivers, we aim to predict water temperature $Y = \{y_{d,t}\}$. In particular, $y_{d,t}$ denotes the temperature at depth $d$ and at time step $t$. 

% The law of conservation of energy states that the difference between incoming and outgoing heat fluxes should result in an equal change of lake energy per unit of time. We model the lake energy at each time step $t$ as $U_t$. We will discuss the computation of heat fluxes and how we build the connection with the change of energy $\Delta U_t$. 

% (Jared) Do we need to say why we chose these drivers?

\section{Preliminaries}
\subsection{General Lake Model (GLM)}

\begin{figure} [!h]
\centering
\includegraphics[width=0.72\columnwidth]{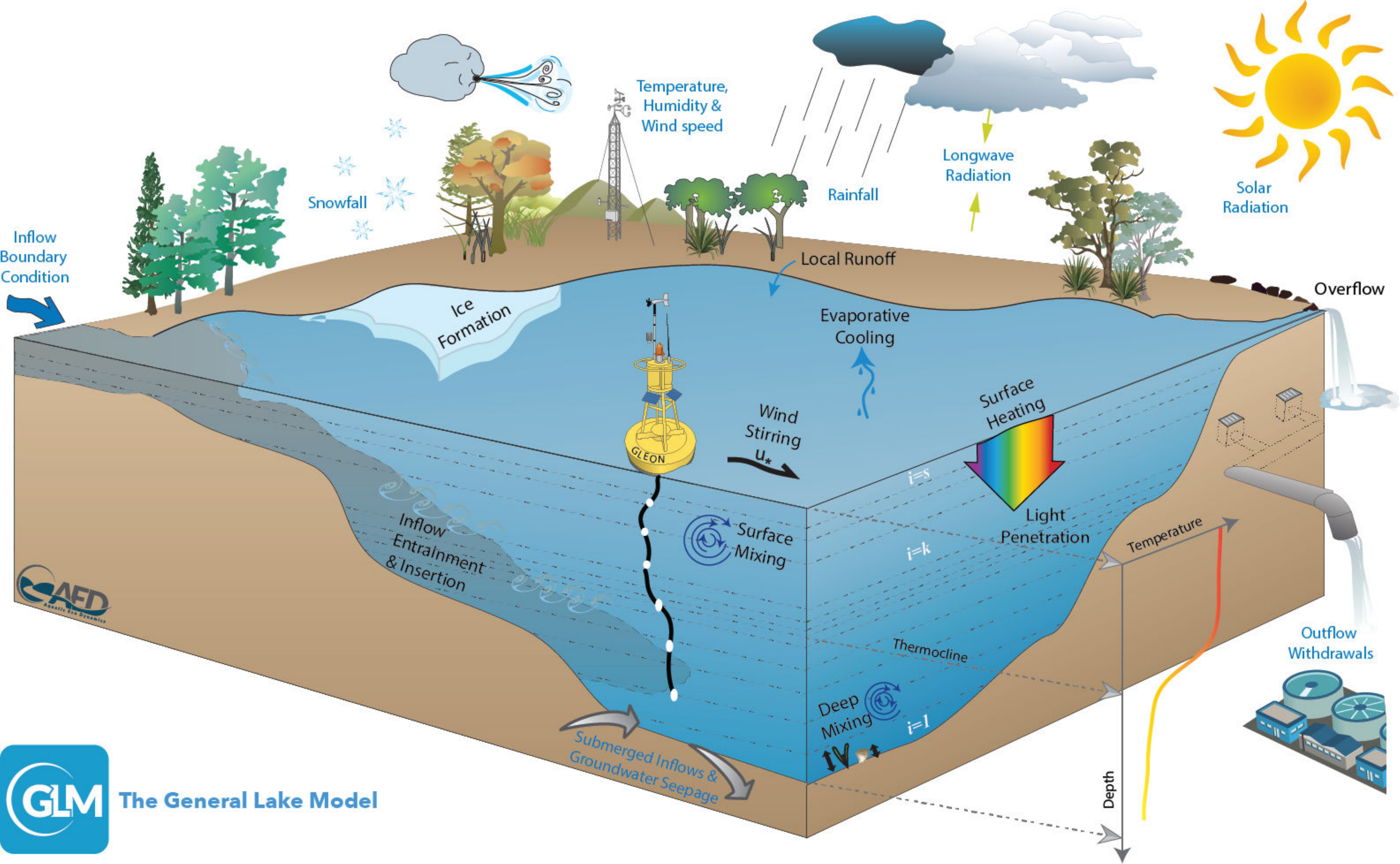}
\caption{A pictorial description of the physical processes simulated by the General Lake Model~\cite{hipsey2014glm}. These processes govern the dynamics of temperature in a lake.}
\label{GLM}
\end{figure}

The physics-based GLM captures a variety of physical processes governing
the dynamics of water temperature in a lake, including the heating
of the water surface due to incoming short-wave radiation, the attenuation of radiation beneath the water surface, the mixing of layers with varying energies
at different depths, and the loss of heat from
the surface of the lake via evaporation or long-wave
radiation (shown in Fig.~\ref{GLM}). We use GLM
as our preferred choice of physics-based model for lake
temperature modeling due to its model performance and wide use among the lake modeling community.

The GLM has a number of parameters (e.g., parameters related to vertical mixing, wind sheltering, and water clarity) that are often calibrated specifically to individual lakes if training data are available. The basic calibration method is to run the model for combinations of parameter values and select the parameter set that minimizes model error. This calibration process can be both labor- and computationally-intensive, and if executed without expert knowledge of parameter meaning and acceptable values, can create model formulations that perform poorly when evaluated against test data. Furthermore, calibration processes applied even in the ideal case of complete and error-free observational are limited by simplifications and rigid formulations of parameters in these physics-based models. 

\subsection{Long-Short Term Memory Networks}
%We first briefly describe the structure of the Long-Short Term Memory (LSTM) model. Given the input  $x_t$ at every time step, t
%
%The LSTM model generates hidden representation $h_t$ at every time step, which are then used for prediction. 
In essence, the LSTM model defines a transition relationship for hidden representation $h_{{t}}$ %at each time step 
through an LSTM cell, %. This LSTM cell 
which combines the input features $x_{t}$ at each time step and the inherited information from previous time steps.

Each LSTM cell contains a cell state $c_t$, which serves as a memory and allows %the hidden units $h_t$ to reserve
reserving information from the past. %The cell state $c_t$ is generated by combining $c_{t-1}$, $h_{t-1}$, and the input features at $t$. Hence, the 
%The transition of cell state over time forms a memory flow, which enables the modeling of long-term dependencies. 
Specifically, the LSTM first generates a candidate cell state $\tilde{c}_t$ by combining $x_t$ and $h_{t-1}$, as: %into a $\text{tanh}(\cdot)$ function:
\begin{equation}
\footnotesize
\begin{aligned}
\tilde{c}_t &= \text{tanh}(W^c_h h_{t-1} + W^c_x x_t).
\end{aligned}
\end{equation}
%where $W^c_h$ and $W^c_x$ are weight parameters used to generate candidate cell state. 
%Hereinafter we omit the bias terms as they can be absorbed into weight matrices. 
%Here the input gate layer is used to determine what information should be kept from the candidate cell state.

Then the LSTM generates a forget gate $f_t$, an input gate  $g_t$, and an output gate  via sigmoid function, as:
%\vspace{-.07in}
\begin{equation}
\footnotesize
\begin{aligned}
%f_t = \sigma(W^f_h h_{t-1} + W^f_x x^t).
f_t &= \sigma(W^f_h h_{t-1} + W^f_x x_t),\\
g_t &= \sigma(W^g_h h_{t-1} + W^g_x x_t),\\
o_t &= \sigma(W^o_h h_{t-1} + W^o_x x_t).
\end{aligned}
\end{equation}

%where $W^f_h, W^f_x,W^g_h, W^g_x$ denote two sets of weight parameters for generating forget gate layer $f^t$ and input gate layer $g^t$, respectively.
%$\sigma(\cdot)$ denotes the sigmoid function, and therefore each entry in the forget/input gate layer ranges in [0,1].
%The forget gate layer is used to filter the information inherited from $c^{t-1}$, and the input gate layer is used to filter the candidate cell state at time $t$. 
The forget gate  is used to filter the information inherited from $c^{t-1}$, and the input gate  is used to filter the candidate cell state at $t$. Then we compute the new cell state and the hidden representation as: %generate the hidden representation $h^t$ at $t$ by filtering the the obtained cell state using the output gate layer $o^t$. This can be summarized as: %In this way we obtain the new cell state $c^t$ as follows:%representing the probability to keep the previous cell state at $i^{th}$ position.
%Based on the $\tilde{c}^t$ and $i^t$, we combine with the cell state from previous time step, and obtain the new cell state as:
\begin{equation}
\footnotesize
\begin{aligned}
c_t &= f_t\otimes c_{t-1}+g_t\otimes\tilde{c}_t,\\
h_t &= o_t\otimes \text{tanh}(c_t),
\end{aligned}
\end{equation}
where $\otimes$ denotes the entry-wise product.
%\yell{change the notation to be subscript}

% According to the above equations, we can observe that the computation of $h_t$ not only depends on $x_{t}$, but also utilizes the inherited information from previous time step (i.e., $h_{t-1}$ and $c_{t-1}$). %In following sections, we represent this process as $h_{t} = \text{LSTM} (x_t| \mathcal{I}_{t-1})$, where $\mathcal{I}_{t-1}$ represents the inherited information from $t-1$.

As we wish to conduct regression for continuous values, we generate the predicted temperature $\hat{y_t}$ at each time step $t$ via a linear combination of hidden units, as:
\begin{equation}
\footnotesize
    \hat{y_t} = W_y h_t.
\end{equation}
%where $W_y$ is the weight matrix to transform hidden representation.

We apply the LSTM model for each depth separately. Given the true observation $y_{d,t}$ at every time step and at every depth, our training loss is defined as:
\begin{equation}
\footnotesize
    \mathcal{L}_{\text{RNN}} = \frac{1}{N_d T}\sum_t \sum_{d} (y_{d,t}-\hat{y}_{d,t})^2,
\end{equation}
where $N_d$ is the total number of different depths, and $T$ is the number of time steps.

\begin{figure} [!h]
\centering
%\raggedleft
\label{fig:b}{}
\includegraphics[width=0.99\columnwidth]{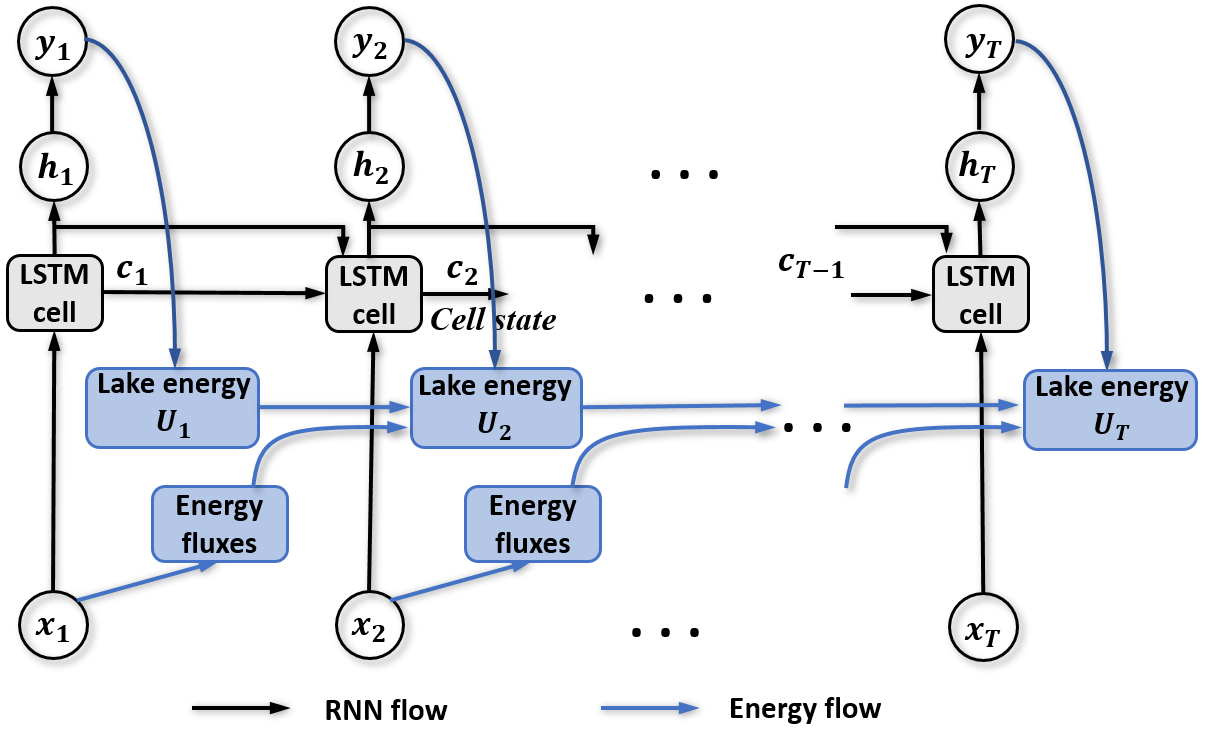}
\caption{The flow of the PGRNN model.} %The model includes the standard RNN flow and the energy flow in the recurrent process.}
\label{flow}
\end{figure}

\section{Method}
\subsection{Energy conservation over time}

We integrate energy conservation flow into the recurrent process, as shown in Fig.~\ref{flow}. While the recurrent flow in the standard RNN can capture data dependencies across time, the modeling of energy flow ensures that the change of lake environment and predicted temperature conforms to the law of energy conservation.

Traditional LSTM models utilize the LSTM cell to implicitly encode useful information at each time step and pass it to the next time step. In contrast, the energy flow in PGRNN explicitly captures the key factor that leads to temperature change in dynamical systems - the heat energy fluxes that are transferred from one time to the next. In this way, the PGRNN provides a structure within to predict what causes the temporal variation in dynamical systems.  Further, note that even though different lake systems have different data distributions, they all conform to the same universal law of energy conservation. Therefore, by complying with the universal law of energy conservation, PGRNN has a better chance at learning generalizable patterns that apply across multiple types of lakes.

%\yell{add more computer science innovation}

\begin{figure} [!h]
\centering
%\raggedleft
\label{fig:b}{}
\includegraphics[width=0.5\columnwidth]{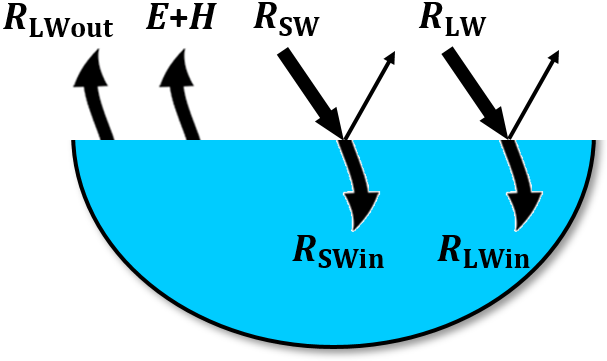}
\caption{The heat energy fluxes that are modeled in PGRNN. For short-wave radiation ($R_{\text{SW}}$) and long-wave radiation ($R_{\text{LW}}$), a portion of the energy is reflected by the lake surface.}
\label{energy}
\end{figure}

%Specifically, at each time step, we compute the heat energy fluxes using the input drivers. 
In Fig.~\ref{energy}, we show the major incoming and outgoing heat fluxes that impact the lake energy. The incoming heat fluxes include terrestrial long-wave radiation and incoming short-wave radiation. The lake loses heat mainly through the outward fluxes of back radiation ($R_{
\text{LWout}}$), sensible heat fluxes ($H$), and latent evaporative heat fluxes ($E$).   %while the outgoing heat fluxes include outward fluxes of back radiation, sensible heat fluxes and latent heat fluxes (evaporation). 
In this work, we ignore the smaller flux terms such as sediment heat flux and advected energy from surface inflows and groundwater.

The balance of these components results in a change in the thermal energy ($U_t$) of the lake. %The change of lake energy $\Delta U_t$ across time should be consistent with the gap between incoming energy fluxes and heat losses, which can be expressed as:
Specifically, the lake will gain energy if the incoming heat fluxes are more than outgoing heat fluxes, and the lake will lose energy if there are more outgoing heat fluxes than incoming heat fluxes. The consistency between lake energy $U_t$ and energy fluxes can be expressed as:
\begin{equation}
\small
\begin{aligned}
    \Delta U_t &= R_{\text{SW}} (1-\alpha_{\text{SW}})+R_{\text{LWin}} (1-\alpha_{\text{LW}})\\
    &-R_{\text{LWout}}-E-H,
\end{aligned}
\label{law}
\end{equation}
where $\Delta U_t=U_{t+1}-U_t$, $\alpha_{\text{SW}}$ is the short-wave albedo (the fraction of short-wave energy reflected by the lake surface) and $\alpha_{\text{LW}}$ is the long-wave albedo. In our implementation, we set $\alpha_{\text{SW}}$ to 0.07 and $\alpha_{\text{LW}}$ to 0.03 which are generally accepted values for lakes from previous scientific studies~\cite{hipsey2017general}.  %(\yell{add citation here}). 
All energy components are in Wm$^{-2}$.

% \yell{mention that this is only applied to non-frozen period.}

Lake Mendota experiences ice cover during winter months, but here we consider the energy conservation only for ice-free periods since the lake exhibits drastically different reflectance and energy loss dynamics when covered in ice and snow, and the modeling of ice and snow was considered out of scope for this study. As such, we define the loss function term for energy conservation and combine this with the training objective of standard LSTM model in the following equation. 
\begin{equation}
\footnotesize
\begin{aligned}
    \mathcal{L} &= \mathcal{L}_{\text{RNN}}+\lambda_{\text{EC}}\mathcal{L}_{\text{EC}},\\
    \mathcal{L}_{\text{EC}}&=\frac{1}{T_{\text{ice-free}}} \sum_{t\in \text{ice-free}} \text{ReLU}(|\Delta U_t-\mathcal{F}|-\tau_{\text{EC}}),
\end{aligned}
\label{combined_1}
\end{equation}
where $\mathcal{F}$ represents the sum of heat fluxes at the right hand side of Eq.~\ref{law}, $\tau_{\text{EC}}$ is the threshold for the loss of energy conservation. This threshold is introduced because physical processes can be affected by unknown less important factors which are not included in the model, or by observation errors in the metereological data. The function $\text{ReLU}(\cdot)$ is adopted such that only the difference larger than the threshold is counted towards the penalty. In our implementation, the threshold is set as the largest value of $|\Delta U_t-\mathcal{F}|$ in the GLM model for daily averages.  %as 24 in our implementation. %\yell{add explanation to threshold and ReLU.} 
The hyper-parameter $\lambda_{\text{EC}}$ controls the balance between the loss of the standard RNN and the energy conservation loss. 

\vspace{.1in}
\noindent\textbf{Estimation of Heat Fluxes and Lake Thermal Energy:} 
%We now introduce how to estimate energy fluxes in our implementation.
% begin introducing each flux component
Terrestrial long-wave ($R_{\text{LWin}}$) radiation is emitted from the atmosphere, and depends on prevailing local conditions like air temperature and cloud cover. Incoming short-wave radiation ($R_{SW}$) is affected mainly by latitude (solar angle), time of year, and cloud cover. Both factors are included in the input drivers $X$.

As for the outgoing energy fluxes, we estimate $E$, $H$, and $R_{\text{LWout}}$ separately using the input drivers and modeled surface temperature.  

The sensible heat flux and latent evaporative heat flux can be computed based on the previous study~\cite{hipsey2017general}: %, as:
\begin{equation}
\footnotesize
\begin{aligned}
    E &= -\rho_a C_E \nu \kappa_{10} \frac{\omega}{p} (e_s-e_a),\\
    H &= -\rho_a c_a C_H \kappa_{10} (T_s-T_a),
\end{aligned}
\label{eh}
\end{equation}
where $C_H$ is the bulk aerodynamic coefficients for sensible heat transfer, and $C_E$ the bulk aerodynamic coefficients for latent heat transfer. Both coefficients are estimated from Hicks' collection of ocean and lake data~\cite{hicks1972some}. The coefficient $\omega$ is the ratio of molecular mass of water to molecular mass of dry air (=0.622), $\nu$ the latent heat of vaporization (=2.453$\times$10$^6$), and $c_a$ the specific heat capacity of air (=1005). 
The variable $T_a$ is the air temperature, and $\kappa_{10}$ the wind speed (m/s) above the lake referenced to 10m height. Both these variables are included or can be derived from input drivers. $T_s$ is the surface water temperature in degrees Kelvin obtained through the feed-forward process. The air density $\rho_a$ is computed as $\rho_a = 0.348 (1+r) /(1+1.61 r) p/T_a$, where $p$ is air pressure (hPa) and $r$ is the water vapour mixing ratio (both derived from input drivers). The vapour pressure ($e_s$ and $e_a$) is calculated by the linear formula from Tabata~\cite{tabata1973simple}:
\begin{equation}
\small
    \begin{aligned}
    e_s &= 10^{(9.28603523\frac{2322.37885}{T_s+273.15})},\\
    e_a &= (S_{\text{RH}}RH/100)e_s,
    \end{aligned}
    \label{esa}
\end{equation}
where $S_{\text{RH}}$ is the relative humidity scaling factor (=1, obtained through calibrating the GLM model) and $RH$ is the relative humidity (included in input drivers).

The back radiation  $R_{\text{LWout}}$ is estimated as: 
\begin{equation}
\footnotesize
    \begin{aligned}
    &R_{\text{LWout}} = \epsilon_s \delta T_s^4,\\
    \end{aligned}
    \label{rout}
\end{equation}
where $\epsilon_s$ is the emissivity of the water surface (=0.97), and $\delta$ is the Stefan-Boltzmann constant (=5.6697e-8 Wm$^{-2}$K$^{-4}$). %, and $T_s$ is the water surface temperature in degrees Kelvin. We use a value of 0.97 for $\epsilon_s$.

% The temperature change in lake water is caused by the energy flow over time. The lake energy budget is a balance between incoming energy fluxes and heat losses from the lake. %A mismatch in losses and gains results in a temperature change - more gains than losses will warm the lake, and more losses than gains will cool the lake. 

%Given the temporal modeling structure in the LSTM model, we add constraint on the predicted temperature over time such that the change of volume-average temperature is consistent to the energy gain/loss. 

On the left-hand side of Eq.~\ref{law}, the total thermal energy of the lake at time $t$ is:
\begin{equation}
\footnotesize
    U_t = c_w \sum_d a_d y_{d,t} \rho_{d,t} \partial{z_d},
    \label{u}
\end{equation}
where $c_w$ is the specific heat of water (4186 J kg$^{-1}$\textdegree{}C$^{-1}$), $a_d$ the cross-sectional area of the water column ($m^2$), and $\partial{z_d}$ the thickness ($m$) of the layer at depth $d$. In this work, we simulate water temperature for every 0.5m and thus we set $\partial{z_d}$=0.5. The computation of $U_t$ requires the output of temperature $y_{d,t}$ through feed-forward process for all the depths, as well as cross-sectional depth area $a_d$.

Note that the modeling of energy flow using the procedure described above does not require any input of true labels/observations. According to Eqs.~\ref{eh}-\ref{u}, the heat fluxes and lake energy are computed using only input drivers and predicted temperature. In light of these observations, we can extend this model to incorporate the energy conservation for other systems which have only a few labeled data points. This enforces predicted temperatures in unmonitored systems to also follow the universal law of energy conservation.

% Note that the heat fluxes and lake energy are computed using only input drivers and predicted temperature (Eqs.~\ref{eh}-\ref{u}), %the modeling of energy flow 
% and does not require any input of true labels/observations. %According to , . 
% In light of these observations, we can extend this model to incorporate the energy conservation for other systems which have only a few labeled data points. This enforces predicted temperatures in unmonitored systems to also follow the universal law of energy conservation. 

\subsection{Pre-training using Physical Simulations}

% In scientific discovery problems, it is critical to build generalizable models that can be applied to different systems. For example, a model built on one lake should be able to generalize to another lake. However, traditional data science models are known to suffer from the data heterogeneity problem where the training data and testing data are from different distributions~\cite{}. In this section, we show that our proposed PGRNN model can improve the generalizability in two ways.  

In many environmental systems, observed data is limited due to the extensive labor required to deploy sensors and process the data. Therefore, the learning model needs to be effectively trained using only small amount of observed data.
For scientific problems, it is also critical to build generalizable models that can be applied to different systems. For example, a model built on one lake should be able to generalize to another lake which has limited observed data. However, traditional data science models are known to suffer from the data heterogeneity problem where training data and testing data are from different distributions~\cite{pan2010survey}. 
Hence, the RNN-based model trained with limited observed data is very likely to lead to unsatisfactory performance. 
% In this section, we show two ways in which our proposed PGRNN model can improve the learning performance using only a few observed data points, and also how we can improve the generalizability.  

% \vspace{.1in}
% \noindent\textbf{Universal law of energy conservation: }
% First, the incorporation of energy flow in PGRNN enables the inclusion of the universal law of conservation of energy in the model. Even though different lake systems have different data distributions, they should all conform to the law of energy conservation. Therefore, the incorporation of energy flow can capture universal temporal patterns for different lake systems.   

% Importantly, %it is noteworthy that 
% the heat fluxes and lake energy are computed using only input drivers and predicted temperature (Eqs.~\ref{eh}-\ref{u}), %the modeling of energy flow 
% and does not require any input of true labels/observations. %According to , . 
% In light of these observations, we can extend this model to incorporate the energy conservation for other systems which have only a few labeled data points. This enforces predicted temperatures in unmonitored systems to also follow the universal law of energy conservation. 

% \vspace{.1in}
% \noindent\textbf{Pre-training using simulated data: }

%In many real-world systems, we may have only a few observed data points due to the substantial manual labor required to employ sensors to measure target variables and the involved data acquisition errors. %involved in this process. 

To address the issue of limited observed data, We pretrain the PGRNN model using the simulated data produced by a simple physics-based model. %This pre-training method aims to  %can significantly improve model performance using only limited observed data. 
In particular, given the input drivers, we can run the GLM to predict temperature at every depth and at every day. These simulated temperature data by GLM %are reasonably accurate and 
conform to underlying physical laws used to build the GLM. Hence, the pre-training using these simulated data can result in a %more accurate and 
physically consistent initialized %status for the learning 
model. When applying the pretrained model to a new system, we fine-tune the model using limited observed data. In our experiments, we show that this pre-training method can achieve high accuracy given very few observed data. 

% The proposed model can be pretrained using simulated data produced by the GLM model. 

\subsection{Density-depth Constraint}
While the modeling of energy flow allows the incorporation of the most important physical process that controls water temperature dynamics, other physical constraints relevant to lake temperature exist. The incorporation of these additional constraints can help guide the model to make predictions that are consistent with real-world physics. To demonstrate the capacity of the PGRNN model to incorporate these constraints, we consider an illustrative example for adding density-depth constraint as follows. 
% While the modeling of lake energy flow takes into account of the progression of overall energy balance with respect to the entire lake, the density-depth constraint provides insight into the relationship between the density of water in different depth layers.  %predictions are physically meaningful. To show   %For example,   

%Having described the hybrid model, we now add additional constraints for training this model so that the predictions are physically consistent. %To better illustrate this, we consider the example of lake temperature monitoring. %We introduce two constraints along the depth dimension and the time dimension, respectively. 
% \vspace{.1in}
% \noindent\textbf{Density-depth constraint: }
It is known that the density of water monotonically increases with depth and thus can be used as constraints on the outputs of PGRNN. We first transform the predicted temperature $Y$ into the density values $\rho$ according to the following known physical equation~\cite{martin2018hydrodynamics}:
\begin{equation}
\footnotesize
    \rho = 1000\times (1-\frac{(Y+288.9414)\times (Y-3.9863)^2}{508929.2\times (Y+68.12963)}).
\label{td}
\end{equation}

%We first transform the values of predicted temperature into the density values according to Eq.~\ref{td}. 
Then, we add an extra penalty for violation of density-depth relationship. Specifically, on any pair of consecutive depths $d$ and $d+1$, if  $\rho_{d,t}$ is larger than $\rho_{d+1,t}$,  then this is considered as a violation to the density-depth relation. In this way, we define the loss of density-depth constraint as:
\begin{equation}
\footnotesize
    \begin{aligned}
    \Delta \rho_{d,t} &= \rho_{d,t}-\rho_{d+1,t},\\
    \mathcal{L}_{\text{DC}} &= \frac{1}{T(N_d-1)}\sum_t \sum_d \text{ReLU} (\Delta \rho_{d,t}),
    \end{aligned}
\end{equation}
where  $\text{ReLU}(\cdot)$  is used to ensure that only pairs with inverse density values are counted towards the penalty. 

% \begin{figure} [!h]
% \centering
% %\raggedleft
% \label{fig:b}{}
% \includegraphics[width=0.4\columnwidth]{depth_density.png}
% \caption{The relation between water density and depth.}
% \label{density}
% \end{figure}

Combining this with Eq.~\ref{combined_1}, the complete training objective becomes:
\begin{equation}
\footnotesize
    \mathcal{L} = \mathcal{L}_{\text{RNN}}+\lambda_{\text{EC}}\mathcal{L}_{\text{EC}}+\lambda_{\text{DC}}\mathcal{L}_{\text{DC}},
\end{equation}
where $\lambda_{\text{DC}}$ is the hyper-parameter to control the penalty for violating the density-depth constraint.

% \begin{table*}[!t]
% \small
% \newcommand{\tabincell}[2]{\begin{tabular}{@{}#1@{}}#2\end{tabular}}
% \centering
% \caption{Performance of RNN, PGRNN, as well as pretrained models from Lake Mendota (RNN$_p$ and PGRNN$_p$) and Florida (RNN$_{\tilde{p}}$ and PGRNN$_{\tilde{p}}$) using different amount of observed data. }
% \begin{tabular}{l|ccccccc}
% \hline
% \textbf{Method} & \textbf{2\%} & \textbf{10\%} & \textbf{20\%} & \textbf{40\%} & \textbf{60\%} &\textbf{80\%} & \textbf{100\%} \\ \hline 
% RNN & 2.311($\pm$0.240) &  1.560($\pm$0.131) &   1.531($\pm$0.083) &  1.525($\pm$0.104) & 1.496($\pm$0.062)  & 1.495($\pm$0.137) &  1.489($\pm$0.091) \\ \hline 
% PGRNN & 2.156($\pm$0.178) &  1.511($\pm$0.135) & 1.484($\pm$0.102) &  1.482($\pm$0.082) &  1.471($\pm$0.093) &  1.470($\pm$0.081) &  1.466($\pm$0.063) \\ \hline 
% RNN$_p$ &  1.650($\pm$0.169)   &  1.487($\pm$0.095)  &   1.417($\pm$0.113)  &  1.401($\pm$0.110)    &  1.397($\pm$0.103) &  1.388($\pm$0.098) &   1.385($\pm$0.080)     \\ \hline 
% PGRNN$_p$ & 1.592($\pm$0.175)   &  1.490($\pm$0.098)  &   1.409($\pm$0.103)  &  1.397($\pm$0.078)  &    1.377($\pm$0.078) &  1.371($\pm$0.094)    &1.377($\pm$0.076)\\ \hline 
% RNN$_{\tilde{p}}$ & 1.930($\pm$0.173) &  1.511($\pm$0.126)  &  1.476($\pm$0.096)  &  1.442($\pm$0.082) &  1.439($\pm$0.075) &   1.438($\pm$0.075)   &    1.398($\pm$0.069) \\ \hline 
% PGRNN$_{\tilde{p}}$ & 1.759($\pm$0.147) &  1.494($\pm$0.108) &  1.470($\pm$0.091) &   1.444($\pm$0.090) &  1.432($\pm$0.054) &   1.426($\pm$0.072) &    1.394($\pm$0.071) \\ \hline 
% \end{tabular}
% \label{general}
% \end{table*}

\section{Experiment}

%We first show how energy conservation helps with prediction and generalization. In addition, we show that PGRNN can result in greater consistency between heat fluxes and lake energy change. Finally, we demonstrate that the incorporation of density-depth constraints leads to more physically consistent predictions. 

%\yell{describe dataset here, add the figure suggested by Vipin}
Our dataset was collected from Lake Mendota in Wisconsin, USA. This lake system is reasonably large ($\sim$40 km$^2$ in area) and displays sufficient dynamics in the temperature profiles over time. Observations of lake temperature were collated from a variety of sources, including the North Temperate Lakes Long-Term Ecological Research Program and a web resource that collates data from federal and state agencies, academic monitoring campaigns, and citizen data~\cite{read2017water}. These temperature observations vary in their distribution across depths and time. There are certain days when observations are available at multiple depths while only a few or no observations are available on some other days.

The input drivers that describe prevailing meteorological conditions are available on a continuous daily basis from April 02, 1980 to December 30, 2014. Specifically, we use a set of 10 drivers as input variables, which include short-wave and long-wave radiation, air temperature, relative humidity, wind speed, frozen and snowing indicators, etc. In contrast, observational data for training and testing the models is not uniform, as measurements were made at varying temporal and spatial (depth) resolutions. In total, 13,158 observations were used for the study period.

% \yell{add results for pre-training using northern lake.}

% \yell{describe each method.}

\begin{table}[!h]
\footnotesize
\newcommand{\tabincell}[2]{\begin{tabular}{@{}#1@{}}#2\end{tabular}}
\centering
\caption{Performance of RNN, PGRNN, as well as pretrained models from Lake Mendota (RNN$_p$ and PGRNN$_p$) and Florida (RNN$_{\tilde{p}}$ and PGRNN$_{\tilde{p}}$) using different amounts of observed data. }
\begin{tabular}{l|ccc}
\hline
\textbf{Method} & \textbf{2\%}  & \textbf{20\%} & \textbf{100\%} \\ \hline 
RNN & 2.311($\pm$0.240) &   1.531($\pm$0.083) &   1.489($\pm$0.091) \\ \hline 
PGRNN & 2.156($\pm$0.178) & 1.484($\pm$0.102) &   1.466($\pm$0.063) \\ \hline 
RNN$_p$ &  1.650($\pm$0.169)   &   1.417($\pm$0.113)   &   1.385($\pm$0.080)     \\ \hline 
PGRNN$_p$ & 1.592($\pm$0.175)   &   1.409($\pm$0.103)    &1.377($\pm$0.076)\\ \hline 
RNN$_{\tilde{p}}$ & 1.930($\pm$0.173) &  1.476($\pm$0.096)  &    1.398($\pm$0.069) \\ \hline 
PGRNN$_{\tilde{p}}$ & 1.759($\pm$0.147)&  1.470($\pm$0.091)  &    1.394($\pm$0.071) \\ \hline 
\end{tabular}
\label{general}
\end{table}

% \begin{table}[!h]
% \footnotesize
% \newcommand{\tabincell}[2]{\begin{tabular}{@{}#1@{}}#2\end{tabular}}
% \centering
% \caption{Performance of RNN, PGRNN, as well as pretrained models from Lake Mendota (RNN$_p$ and PGRNN$_p$) and Florida (RNN$_{\tilde{p}}$ and PGRNN$_{\tilde{p}}$) using different amount of observed data. }
% \begin{tabular}{l|ccccc}
% \hline
% \textbf{Method} & \textbf{0\%}& \textbf{0.2\%} & \textbf{2\%}  & \textbf{20\%} & \textbf{100\%} \\ \hline 
% RNN & - & 4.769 & 2.311 &   1.531 &   1.489 \\ \hline 
% PGRNN & - & 4.118 & 2.156 & 1.484 &   1.466 \\ \hline 
% RNN$_p$ & 2.321 & 2.159 & 1.650   &   1.417   &   1.385    \\ \hline 
% PGRNN$_p$ & 2.431 & 2.060 & 1.592   &   1.409    &1.377\\ \hline 
% RNN$_{\tilde{p}}$ & 8.657 & 3.498& 1.930 &  1.476  &    1.398 \\ \hline 
% PGRNN$_{\tilde{p}}$ &9.010 & 2.580& 1.759&  1.470  &    1.394 \\ \hline
% \end{tabular}
% \label{general}
% \end{table}
\subsection{Performance for prediction and generalization}

We use the observed data from  April 02, 1980 to October 31, 1991 and from June 01, 2003 to December 30, 2014 as training data (in total 8,037 observations). Then we apply the trained model to predict the temperature at different depths for the period from November 01, 1991 to  May 31, 2003 (5,121 observations). 

%% compare with GLM
%To get a sense of how well PGRNN performs, 
To give a sense of the overall performance of the many variations of the PGRNN approach we evaluate in this section, we provide the following statistics for the state-of-the-art GLM model. The uncalibrated GLM model achieves the RMSE of 2.950 in the test period. The GLM model can also be fine-tuned to fit each lake system by optimizing the parameter set to minimize model error~\cite{soranno2017lagos}. If we use the same 2\% training data to optimize parameters in the GLM model, it can reach the test RMSE of 2.645. If we fine-tune it using 100\% training data, it will reach the test RMSE of 2.253.

To verify that energy conservation and pre-training helps improve the prediction and genrealization, we compare RNN and PGRNN in terms of their prediction RMSE. Here we do not include the basic neural network since it produces RMSE around 1.8, which is far higher than standard RNN.

To test whether each model can perform well using reduced observed data, we randomly select different proportion of data from the training period. For example, to select 20\% of training data, we remove every observation in our training period with 0.8 probability. The test data stays the same regardless of training data selection. We repeat each test 10 times and report the mean RMSE and standard deviation. %The results are reported in Table~\ref{general}.

% EC improves the prediction
According to Table~\ref{general} (rows 1-2), we can observe that PGRNN consistently outperforms standard RNN. 
The gap is especially obvious when using smaller subsets of observed data (e.g., 2\% data). 
%and PGRNN$_p$ also performs better than RNN$_p$. Therefore, we can conclude that the incorporation of energy conservation improves the prediction accuracy. 
% EC helps with the generalization
%It can be seen that the standard RNN leads to a large error using only limited observed data (e.g., 2\% data). In contrast, 
PGRNN can reach reasonable accuracy using a small amount of observed data because the law of energy conservation regularizes the model to retain physical consistency. 

We can see that PGRNN using only 2\% observed data can outperform the fully calibrated GLM model (using 100\% data). The PGRNN model tuned using 100\% data and the pretrained models can achieve much lower RMSE than the GLM model.

% Moreover, the superiority of PGRNN$_{\tilde{p}}$ over RNN$_{\tilde{p}}$ shows that the modeling of energy conservation can also help better generalize the pretrained model from another lake to Lake Mendota. After incorporating energy conservation, the fine-tuning using only small amount of observed data in Mendota can produce satisfactory result. 

We also verify that the pre-training can indeed improve prediction accuracy and generalizability of the model.  
A basic premise of pre-training our models is that GLM simulations, though imperfect, provide a synthetic realization of physical responses of a lake to a given set of meteorological drivers. Hence, pre-training a neural network using GLM simulations allows the network to emulate a synthetic but physical phenomena. Our hypothesis is that such a pretrained model requires fewer labeled samples to achieve good generalization performance, even if the GLM simulations do not match with the observations. To test this hypothesis, we conduct an experiment where we generate GLM simulations with input drivers from Lake Mendota. These simulations have been created using a GLM model with generic parameter values that are not calibrated for Lake Mendota, resulting in large errors in modeled temperature profiles with respect to the real observations on Lake Mendota (RMSE=2.950). Nevertheless, these simulated data are physically consistent and by using them for pre-training, we can demonstrate the power of our ML models to work with limited observed data while leveraging the physics inherent in the physical models. 

% pre-training improves the prediction and generalization
We pre-train RNN and PGRNN using such simulated data and we report their performance when fine-tuned with true observations in Table~\ref{general} (rows 3-4). %
The comparisons between RNN and RNN$_p$ and between PGRNN and PGRNN$_p$ show that the pre-training can significantly improve the performance. The improvement is especially obvious given small amount of observed data. Moreover, we find that the training RNN and PGRNN model commonly takes 150-200 epochs to converge while the training for RNN$_p$ and PGRNN$_p$ only takes 30-50 epochs to converge. This demonstrates that pre-training can provide a better initialized status for the learning model. 

To assess the generalization ability of ML models in input conditions different from what we have observed previously, we conduct a different experiment. Here we generate GLM simulations for a synthetic lake with input drivers from Florida (which are very different from the typically much colder conditions in Wisconsin) and then pretrain RNN and PGRNN using the simulated data by GLM based on these input drivers. We show the performance of pretrained models (RNN$_{\tilde{p}}$ and PGRNN$_{\tilde{p}}$) in Table~\ref{general} (rows 5-6).

While RNN$_{\tilde{p}}$ and PGRNN$_{\tilde{p}}$ trained using these input drivers and simulated data in Florida have very poor performance when directly applied to Lake Mendota (RMSE=9.010 for RNN$_{\tilde{p}}$ and 8.657 for PGRNN$_{\tilde{p}}$), once they are refined using observed data from Lake Mendota, they get much better.  Here we note that PGRNN$_{\tilde{p}}$ improves with a much greater amount than RNN$_{\tilde{p}}$, which shows the increased generalization power obtained by having a model with physics built into it.

GLM has rarely been tested on more than a handful of years worth of observation data for a single lake. Even when trained and tested on only 2 years of observations, which should produce lower RMSE, GLM model performance is still worse than our PGRNN model performance~\cite{bruce2018multi}. Our PGRNN model performance for lake water temperature over multiple decades is unprecedented and will be a valuable tool for water resource managers and lake scientists.

\begin{figure} [!h]
\centering
%\raggedleft
\subfigure[]{ \label{m1}
\includegraphics[width=0.8\columnwidth]{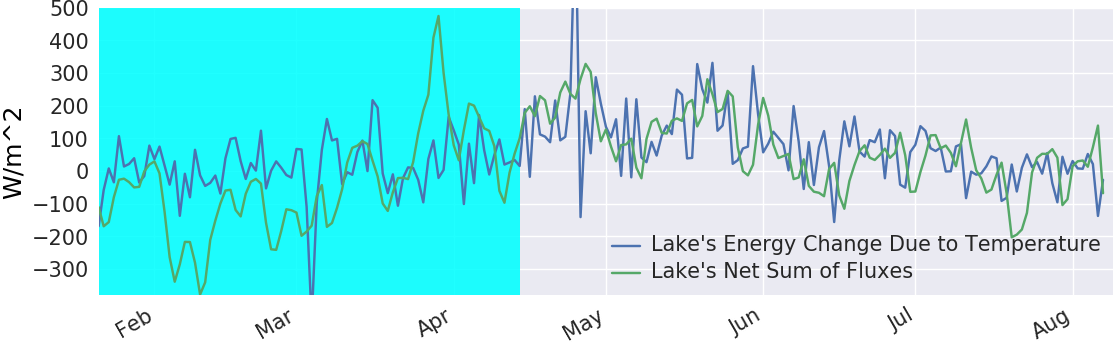}
}\vspace{-.05in}
\subfigure[]{ \label{m2}
\includegraphics[width=0.8\columnwidth]{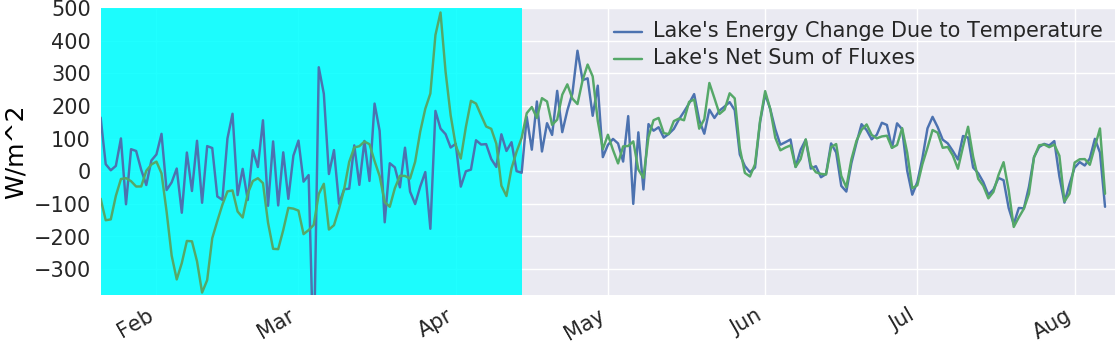}
}\vspace{-.05in}
\subfigure[]{ \label{m2}
\includegraphics[width=0.8\columnwidth]{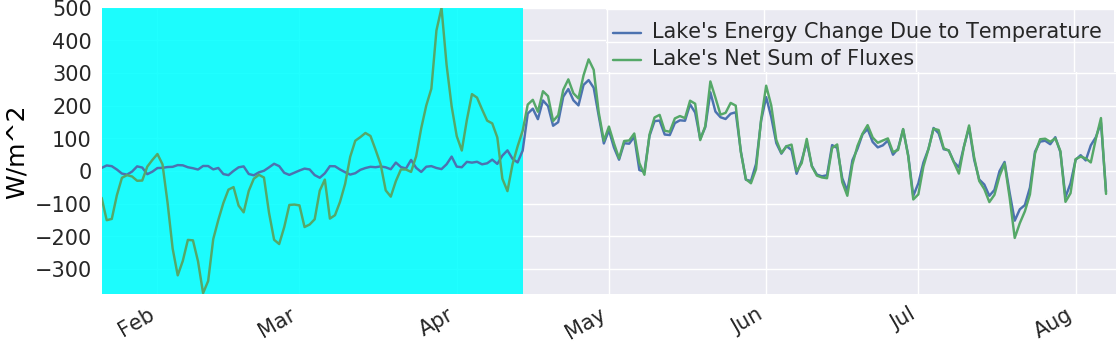}
}\vspace{-.1in}
\caption{The sum of heat fluxes and the lake energy change genereated by (a) RNN, (b) PGRNN, and (c) the generic GLM, from January 21, 1989 to August 09, 1989. The blue part on the left indicates the frozen period (where we do not apply energy conservation).}
\label{fluxes}
\end{figure}

\noindent\textbf{Heat fluxes and lake energy change} 
To visualize how PGRNN contributes to a physically consistent solution, we testify whether the gap between incoming and outgoing heat energy fluxes matches the lake energy change over time. Specifically, we train RNN and PGRNN using observed data from the first ten years. Then, we show the curves for the gap between incoming and outgoing heat fluxes and the change of lake energy over time for a certain test period (Fig.~\ref{fluxes}). These two curves should be well aligned (in the ice-free period) if the learning model follows the energy conservation. % law of energy conservation.  %\yell{train using first ten years.} during some certain periods (Fig.~\ref{fluxes}).

It can be seen that the two curves generated by standard RNN have significant differences with each other, which indicates the physical inconsistency over time. In contrast, by modeling the energy flow, PGRNN can generate energy fluxes that to a greater degree match the lake energy change. Because our energy conservation approach has ignored minor terms (e.g., energy fluxes through sediment heating or advection), it is expected that these curves do not completely overlap. Nevertheless, this close agreement confirms that the PGRNN model is guided by the universal law of energy conservation over time. 
\begin{figure} [!h]
\centering
%\raggedleft
\label{fig:b}{}
\includegraphics[width=0.45\columnwidth]{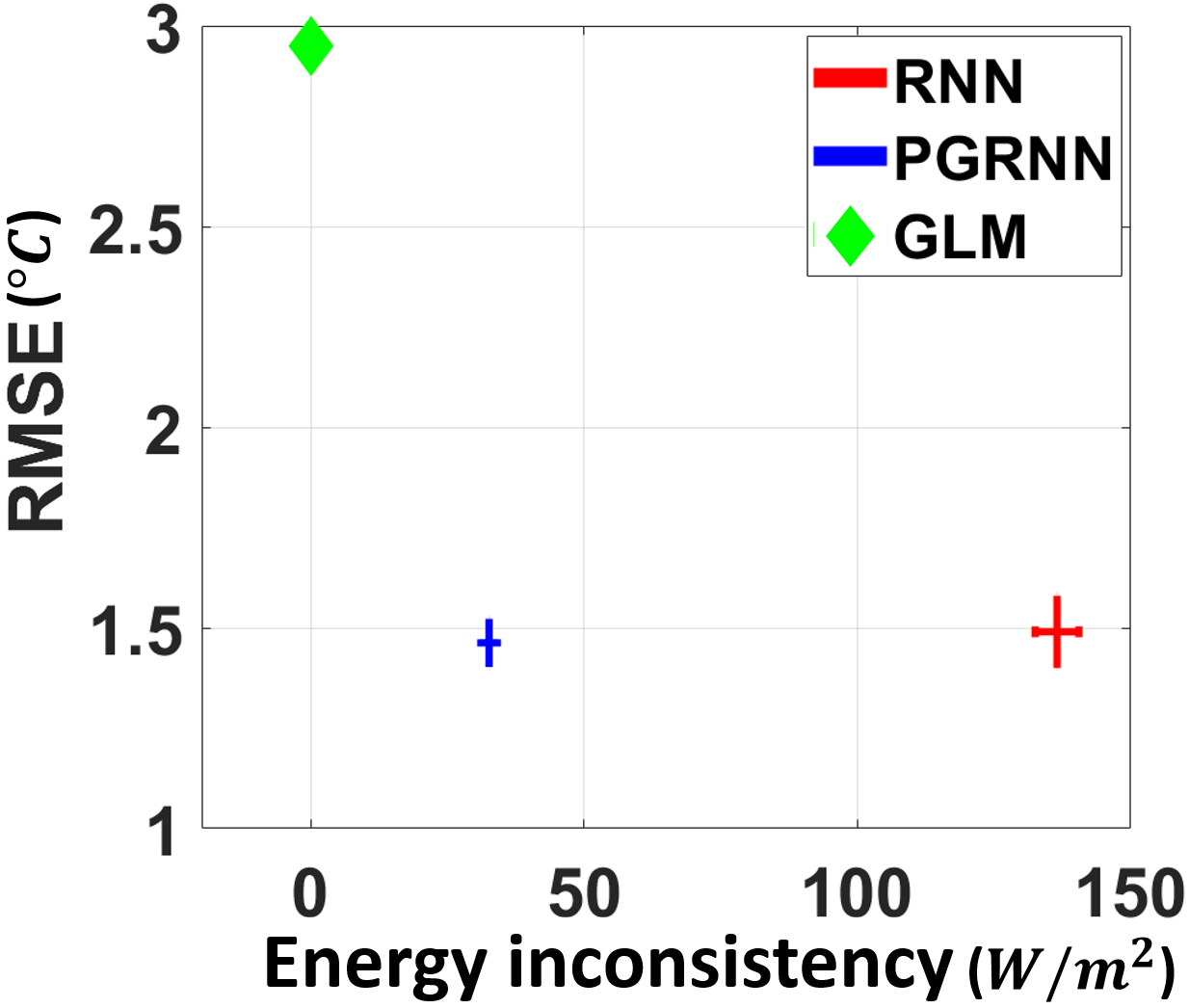}
\caption{The performance of RNN, PGRNN and generic GLM by RMSE and energy inconsistency.}
\label{ec_res}
\end{figure}

% \yell{Explain why Energy inconsistency not going to zero?}

In Fig.~\ref{fluxes} (c), it can be seen that the two curves produced by the generic GLM model are even better aligned than the the curves obtained from the PGRNN model. In real-world systems, the temperature data can be noisy or affected by other unknown factors. Hence, we adopt the soft Lagrangian regularization (Eq.~\ref{combined_1}), which may not lead to the perfect match between heat fluxes and lake energy change.  

We summarize the average gap between these two curves in ice-free periods as the energy inconsistency.  In Fig.~\ref{ec_res}, we show the RMSE and the energy inconsistency of RNN, PGRNN and the generic GLM model in the entire test period. Compared with RNN and GLM, PGRNN can significantly reduce both prediction RMSE and energy inconsistency.

\subsection{Density-depth relation}
We now show that the incorporation of density-depth constraint can further improve the prediction. Specifically,  from all the pairs of consecutive depths, we compute the fraction of pairs where the model makes physically inconsistent predictions (i.e., the density-depth relationship is violated). We report the average of this fraction over all the test data as the measure of density inconsistency. %Note that this measure
%does not require actual observations, and hence,
%we compute this measure over the plentifully large
%unlabeled data set.

\begin{figure} [!t]
\centering
%\raggedleft
\label{fig:b}{}
\includegraphics[width=0.45\columnwidth]{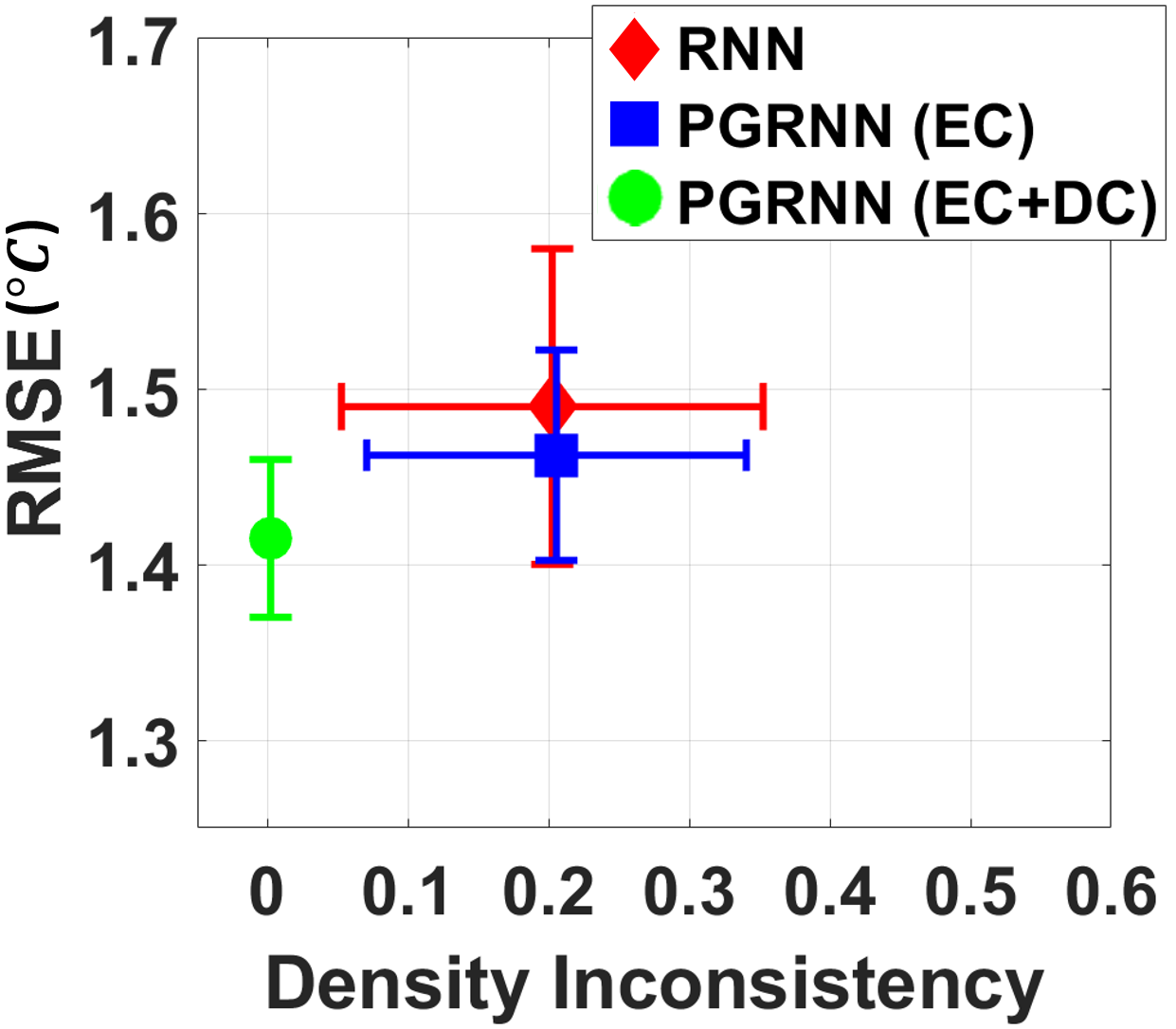}
\caption{The performance of RNN and PGRNN (EC/EC+DC) by RMSE and density inconsistency.}
\label{dc_res}
\end{figure}

According to Fig.~\ref{dc_res}, we can observe that PGRNN reduces RMSE compared with the standard RNN, but still retains high density inconsistency. In contrast, the incorporation of the density-depth constraint effectively reduces density inconsistency to 0.0021. 

\begin{figure} [!h]
\centering
%\raggedleft
\label{fig:b}{}
\includegraphics[width=0.5\columnwidth]{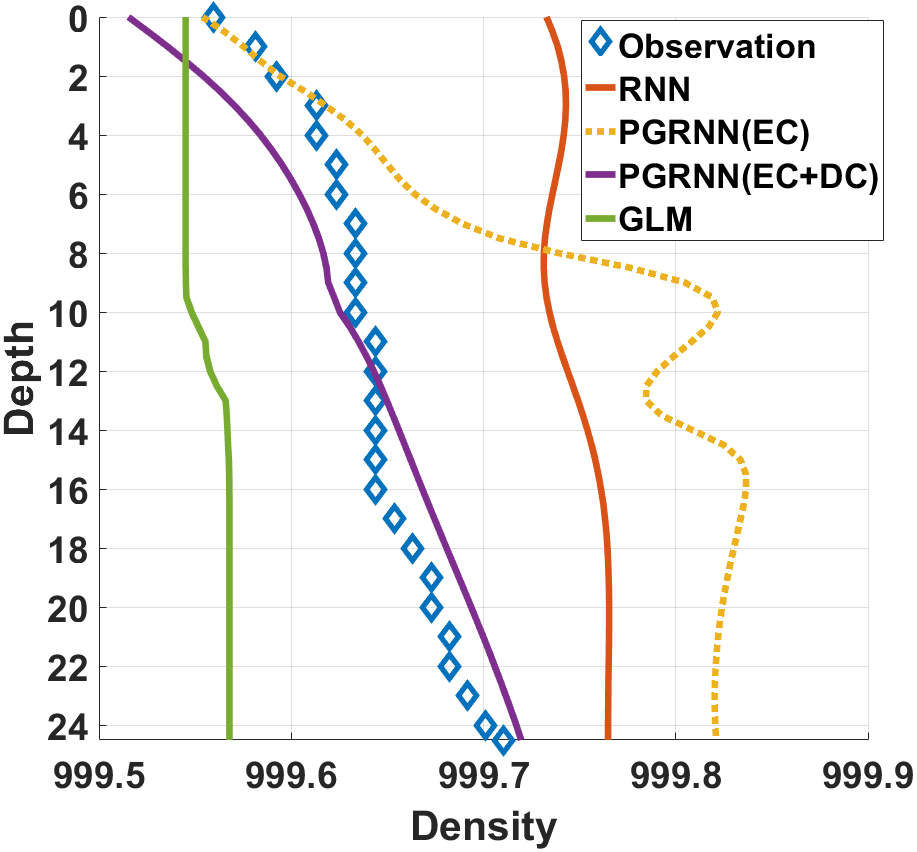}
\vspace{-.07in}
\caption{The obtained density values at different depths by different methods and ground-truth observations on May 20, 2002.}
\label{den_exp}
\end{figure}

In Fig.~\ref{den_exp}, we show the predicted density values (computed from temperature by Eq.~\ref{td}) by RNN, PGRNN and PGRNN+DC on a certain date when observations are available at most depths. It can be seen that the RNN results in a high prediction error since the predictions stay far away from true observations. Besides, the density values start to decrease below depth 15m, which violates the density-depth constraint. Compared with the standard RNN, the PGRNN improves the prediction accuracy, but some predicted values still violate the density-depth constraint (at depth 9-12m). In contrast, the PGRNN+DC model achieves both higher overall accuracy and physically meaningful density predictions.

\subsection{Sensitivity test}
\begin{figure} [!h]
\centering
%\raggedleft
\subfigure[]{ \label{m1}
\includegraphics[width=0.49\columnwidth]{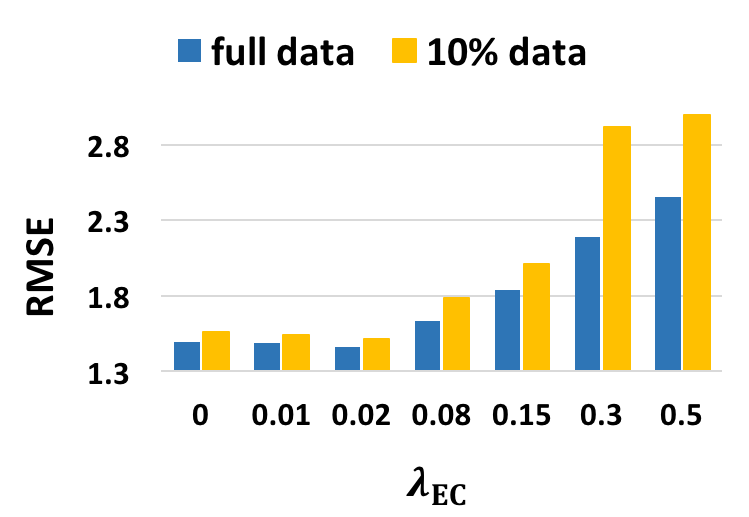}
}\hspace{-.2in}
\subfigure[]{ \label{m2}
\includegraphics[width=0.49\columnwidth]{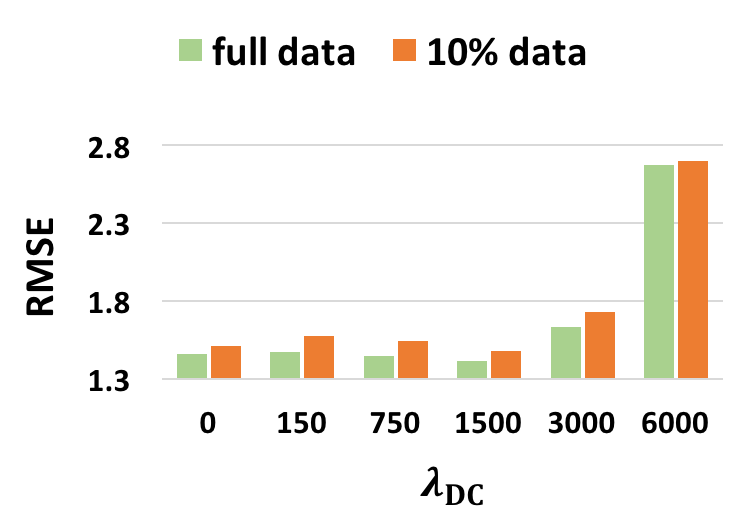}
}\vspace{-.1in}
\caption{Sensitivity tests: the performance (with full data or 10\% data) with respect to (a) the weight of energy conservation $\lambda_{\text{EC}}$, and (b) the weight of density-depth constraint $\lambda_{\text{DC}}$.}
\label{sens}
\end{figure}

We conduct sensitivity tests to examine the impact of hyper-parameters ($\lambda_{\text{EC}}$ and $\lambda_{\text{DC}}$) on the performance (Fig.~\ref{sens}). 

As observed from Fig.~\ref{sens} (a), the performance improves when we increase the weight of energy conservation loss from 0 to 0.02. However, as we adopt extremely large $\lambda_{\text{EC}}$, the RMSE will be high due to the ignorance of the standard supervised training loss. 

In Fig.~\ref{sens} (b), we can see that the performance is relatively stable as we add weight to the density-depth constraint. However, when the value of $\lambda_{\text{DC}}$ is larger than  3000, the error starts to increase sharply. This is also because the training focuses on generating physically meaningful outputs while failing to fit the observed data.

\section{Related Work}
The standard and most common approach for incorporating physical knowledge in machine learning models is to use it to guide feature selection or feature construction, but such approaches cannot learn in absence of labels.  Recently, several papers have considered the use of constraints (based on physical knowledge) on the output of a machine learning model so that these models can be trained even with unlabeled data by relying on physical principles~\cite{ren2018learning,stewart2017label,karpatne2017physics}.  Another very common approach is residual modeling, where an ML model is used to predict the errors made by a physics-based model~\cite{forssell1997combining,hamilton2017hybrid,xu2015data}. In addition to generating results that may be inconsistent with physical laws, these residual modeling approaches do not have the capability to train ML models with unlabeled data by relying on physical principles.  The approach presented in~\cite{karpatne2017physics} is related to both of these types of approaches, as it uses the output of a physics-based model as an additional feature in a neural network model (making it related to residual modeling approaches), and also makes use of constraints on the output to allow for label free learning.  
The PGRNN approach presented in this paper is unique in that it provides a powerful framework for modeling spatial and temporal processes.  This framework makes it possible to incorporate complex physical laws such as  energy and mass conservation (that cannot be enforced by just considering the output of the model) and can also be pre-trained using physics based models.  Together these properties enable development of models that are much more generalizable and can learn with little observational data.

\section{Acknowledgement}
This work was funded by the USGS grant  G18AC00352, NSF Awards 1029711 and DTC seed grant. JSR was funded by the Department of the Interior Northeast and North Central Climate Adaptation Science Centers. Access to computing facilities was provided by  Minnesota Supercomputing Institute.

\section{Conclusion}
In this paper, we propose a novel learning model PGRNN that integrates energy conservation and a density-depth constraint into standard recurrent neural networks for monitoring dynamical systems for scientific knowledge discovery.  For the prediction of lake temperature dynamics, RNN simulations had better performance (RMSE) than a commonly used physics-based model with an optimized parameter set~\cite{bruce2018multi}, but suffered physical inconsistencies such as violations of the conservation of energy and denser water being on top of less dense water. Our PGRNN approach, which used a loss function with physical constraints, reduced or eliminated two different types of physical inconsistencies (energy conservation and depth-density consistency) while also improving the model accuracy.

We also studied the ability of pre-training these models using simulated data to deal with the scarcity of observed data. Using the simulated data from a poorly parameterized physics-based model, we observed an increasing model performance of PGRNN over RNN with fewer observation data used for training.  Thus, PGRNN can leverage the strengths of physics-based models while also filling in knowledge gaps by overlaying features learned from data.

The proposed method can also be adjusted to model other important physical laws in dynamical systems, such as the law of mass conservation. Since energy conservation and mass conservation are universal laws in dynamical systems, the proposed PGRNN model can be applied to a variety of scientific problems such as nutrient exchange in lake systems and analysis of crop field production, as well as engineering problems such as auto-vehicle refueling design. Moreover, the proposed model allows incorporation of additional physical constraints specific to different tasks. Therefore, we anticipate this work to be an important stepping-stone towards more innovations of machine learning for scientific knowledge discovery. 

An extended version of this paper is available at~\cite{jia2019PGRNN_arxiv}.

% In this paper, we propose a learning model PGRNN that integrates energy conservation into standard recurrent neural networks for monitoring dynamical systems for scientific knowledge discovery. The experimental results show that the proposed model can significantly improve the prediction while also maintaining the physical consistency. Moreover, we show that the incorporation of the density-depth constraint in lake temperature monitoring can also help reduce density inconsistency. %relationship. This can help produce more physically meaningful results to specific learning tasks.

% The proposed method can also be adjusted to model other important physical laws in dynamical systems, such as the law of mass conservation. Since energy conservation and mass conservation are universal laws in dynamical systems, the proposed PGRNN model can be applied to a variety of scientific problems such as nutrient exchange in lake systems and analysis of crop field production, as well as engineering problems such as auto-vehicle refueling design. Moreover, the proposed model allows incorporation of additional physical constraints specific to different tasks. Therefore, we anticipate this work as an important stepping stone towards more innovations of machine learning for scientific knowledge discovery.

% \yell{1. GLM energy-based flow can be used to analyze other water properties. 2. propose a generic framework that allows incorporation of more sophisticated domain physics for other scientific research.}

% \section{Acknowledgement}

\renewcommand\refname{Reference}
\bibliographystyle{plain}
\bibliography{reference}

\end{document}